\documentstyle[11pt,IAU207_pasp,twoside,psfig]{article}
\def\edcomment#1{\iffalse\marginpar{\raggedright\sl#1\/}\else\relax\fi}
\reversemarginpar

\begin{document}
\title{The Star Cluster Systems of the Magellanic Clouds}

\author{G. S. Da Costa}

\affil{Research School of Astronomy \& Astrophysics, ANU, Mt.\ Stromlo
Observatory, Cotter Rd, Weston ACT 2611, Australia}

\begin{abstract}
The characteristics of the cluster systems of the Magellanic Clouds, as
inferred from integrated properties, are compared with those from individual
cluster studies and from the field population.  The agreement is generally
satisfactory though in the case of the LMC, the lack of clusters older
than $\sim$3 Gyr is not reflected in the field population.  The possible 
origin(s) for this cluster ``age-gap'' are discussed.  The SMC cluster 
age-metallicity relation is also presented and discussed.
\end{abstract}

\section{Introduction}

The star cluster systems of the Magellanic Clouds play a pivotal role
in fulfilling this Symposium's aim: ``to gain a comprehensive picture of
star cluster formation and evolution, and their role in the evolutionary
framework of their parent galaxies''.  This is because the
relative proximity of the Magellanic Clouds means that LMC and SMC star
clusters can be studied using both {\it integrated light} techniques and
methods that rely on the {\it light of individual cluster stars}.  We can
then investigate the answer to the question ``Do inferences made from 
integrated cluster
properties agree with those from detailed star-by-star analyses?''
Obviously the answer is important for studies of more
distant cluster systems where only integrated light techniques are 
possible.  Similarly, the proximity of the Magellanic Clouds
makes it also possible to study the field star, as distinct from the cluster,
populations of these galaxies. We can seek then to answer
the question ``Do results determined from the characteristics of the 
cluster system, such as `Star' Formation Histories, agree with those for the
general field population?''.  Again the answer is vital if the assumption
that star cluster systems are surrogates for field populations is to 
produce valid results.  

These questions underlie the basic approach of this contribution, in which
we will compare the results of studies of the integrated properties of
the Magellanic Cloud cluster systems, i.e.\ those properties an observer
in a distant galaxy might determine, with the results from
studies of individual LMC and SMC clusters, and of the field star populations.
We begin with the LMC, and then discuss the SMC for which there are generally
fewer data available.  

\section{The Large Magellanic Cloud} 

Figure 1 shows in the left panel, a colour-magnitude (c-m) diagram, and in the 
right panel, a ($U-B$, $B-V$) 2-colour diagram, for LMC star clusters using 
the data of Bica et al.\ (1996).  Our distant observer of the LMC cluster
system would immediately be able to conclude from these data that cluster 
formation has been on-going in the LMC, but to be more quantitative, it is 
necessary to compare the observations with single stellar
population (SSP) models.  In what follows we make use of the models (and 
results) of Girardi et al.\ (1995).  These authors have calculated SSP models
for a variety of abundances and find that they can achieve satisfactory
agreement between the model tracks and the Bica et al.\ (1996) observational
sample for abundances appropriate for the LMC.

\begin{figure}[t]
\centerline{\vbox{
\psfig{figure=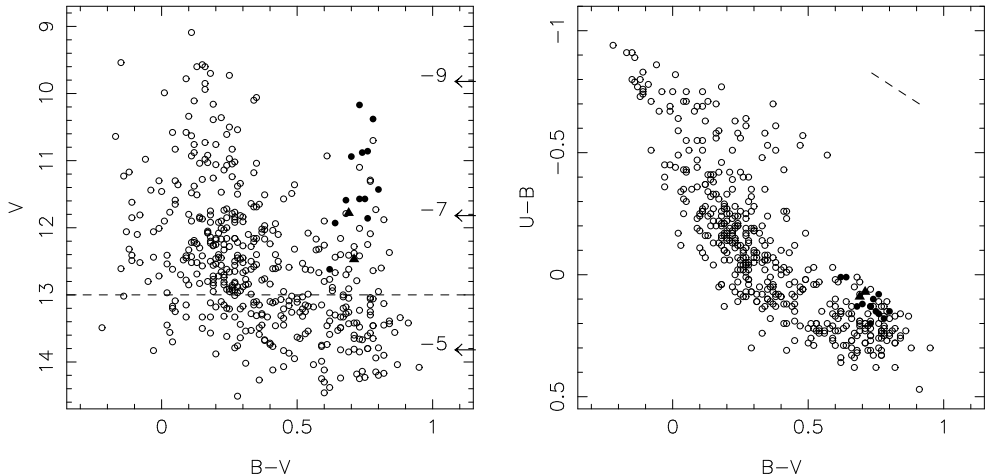,width=13.0cm,angle=270}}}
\caption{Integrated photometry for LMC Star Clusters from Bica et al.\ (1996).
In both panels filled circles are confirmed 
old (age $>$ 10 Gyr) clusters while the filled triangles are the candidate
old clusters from Dutra et al.\ (1999).  Absolute magnitudes for $(m-M)_{0}$
= 18.5 and $E(B-V)$ = 0.10 mag are indicated on the right side of
the left panel.  The horizontal dashed line in the left panel is the 
approximate completeness level.  The dashed line in the right panel shows
the effect of a reddening change of 0.2 mag.}
\end{figure}

\subsection{Clusters younger than $\sim$1 Gyr}

Based on the SSP models of Girardi et al.\ (1995), stars clusters younger
than $\sim$1 Gyr are those with $(B-V)_{0}$ $\leq$ $\sim$0.50 mag.
The {\it observed} age distribution function for these clusters represents the
global cluster formation history modified by the effects of observational
selection (likely to be a minor issue here), fading as clusters age, and
cluster disruption processes.  The fading of clusters with age can be 
compensated for in a straightforward manner via SSP models, but allowance
for the variety of potential cluster disruption processes is more difficult.
Nevertheless, it is reasonable to assume that cluster disruption processes
vary with age in a smooth way, and thus narrow 
features, such as peaks in the observed cluster age distribution at
particular times, can be reasonably interpreted as relative increases 
in the true cluster formation rate at those times.

Girardi et al.\ (1995) determined the observed age distribution function as
follows.  As did Elson \& Fall (1985), they first drew a fiducial curve
through the cluster points in the right panel of Fig.\ 1.  An age dependent
parameter {\it s} is then defined as the distance along this fiducial line.
The relation between {\it s} and age is fixed by determining ages,
using the same theoretical models as are incorporated in the SSP models,
for a set of clusters with well determined c-m diagrams.  
The cluster points are then projected back to the fiducial line and a histogram
distribution of {\it s} values, or equivalently an age distribution, generated.
Girardi et al.\ (1995) differ from earlier work (cf.\ Elson \& Fall) in that
the projection vectors were derived by using the SSP models to calculate,
as a function of cluster age, the effects on the integrated colours of stochastic variations in the numbers of cluster stars.  
These variations reproduce well the scatter seen in the left panel of Fig.\ 1.

The resulting observed cluster age distribution (cf.\ Figs.\ 12 and 15
of Girardi et al.\
1995) has prominent peaks, which represent relative enhancements in the
cluster formation rate, at ages of approximately 10 and 100 Myr.  The
location of these peaks is in {\it remarkedly good agreement} with the peaks
in the observed cluster age distribution given by Pietrzy\'{n}ski \& Udalski
(2000; these proceedings).  These authors have individually age-dated 
approximately 600 LMC star clusters using c-m digrams derived from
the OGLE photometry database.  The peak in the relative number of clusters
at ages around $\sim$100 Myr also has a correspondance in the field star data.
For example, Frogel \& Blanco (1983), among others, have used the colours
and luminosities of LMC red giants in the LMC Bar to infer an era of enhanced
(field) star formation at approximately this epoch.  

Thus for this range of ages, the integrated cluster, the individual cluster,
and the field star studies all give similar results.  

\subsection{Clusters older than $\sim$1 Gyr}

\subsubsection{2.2.1 The (very) old population.}

The LMC has 13 (possibly 15, see Dutra et al.\ 1999) genuine old globular
clusters (e.g.\ Suntzeff et al.\ 1992).  The integrated data for 
these clusters, i.e.\ their colours and spectra, are generally consistent 
with those for the Galactic halo globular clusters (e.g.\ Searle et al.\ 1980), 
supporting a strong degree of similarity.  Indeed, based on c-m diagrams 
derived from HST/WFPC2 images, these LMC clusters appear to have the same age 
as Milky Way globular clusters to within a precision of approximately $\pm$1.5 
Gyr (Olsen et al.\ 1998; Johnson et al.\ 1999; Johnson et al., these
proceedings).  

However, as shown in Fig.\ 1, the LMC analogues of the Galactic
halo globular clusters are not readily distinguished from the numerous
intermediate-age ($\sim$1 $\leq$ age $\leq$ $\sim$10 Gyr) LMC star clusters
in either the ($V, B-V$) or the ($U-B, B-V$) diagrams.  Indeed it would be 
brave (or foolish) for our ``distant observer'' to attempt to identify, using
these data alone, which of the LMC clusters with red ($B-V$) integrated colours 
are the analogues of the Galactic halo globular clusters.  Nevertheless, it is
possible to make some progress in this direction by: (i) restricting the
sample of clusters to the more luminous objects in order to reduce the effect
of stochastic variations (i.e.\ sampling errors) on the integrated colours;
(ii) determining individual reddenings for each cluster; and (iii) adding
observations in near-IR wavebands.  In star
clusters younger than the Galactic halo globular clusters,
the evolution of asymptotic giant branch stars continues to
luminosities above that of the red giant branch tip.  These upper-AGB stars
are cool and so have their greatest influence on near-IR colours.  Fig.~2
shows ($(U-B)_{0}$, $(B-V)_{0}$) and ($(J-K)_{0}$, ($B-V)_{0}$) 2-colour
diagrams for a restricted set of LMC star clusters using the data from
Persson et al.\ (1983).  In these diagrams the genuine old LMC star clusters
can be separated from the intermediate-age clusters, as the latter have
redder $(J-K)_{0}$ and $(U-B)_{0}$ for a given $(B-V)_{0}$\footnote{The
clear separation of the old and intermediate-age clusters in Fig.\ 2
may well be due to the lack of clusters in the LMC with ages between $\sim$3 
and 10 Gyr.}.  As regards our distant observer then, these data allow 
the conclusion that there is a significant age range amongst the ``red'' LMC
star clusters.

\begin{figure}[t]
\centerline{\vbox{
\psfig{figure=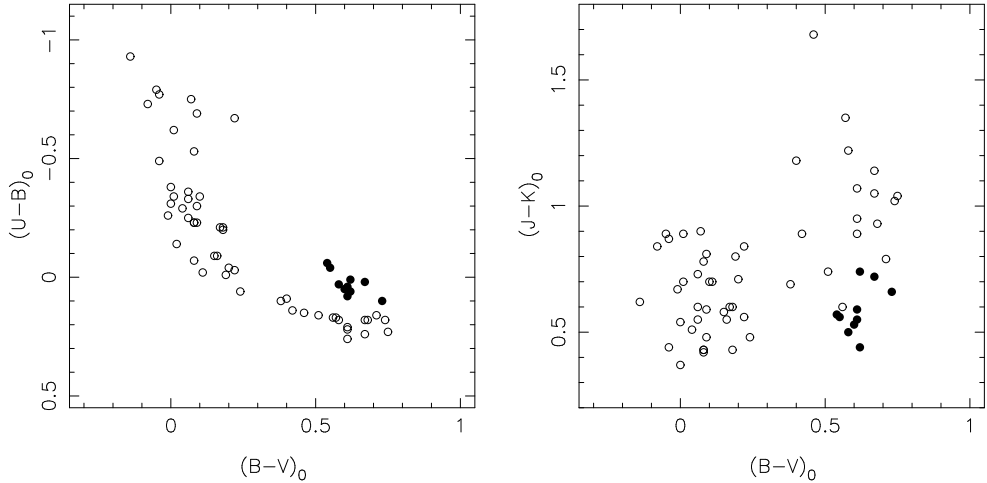,width=13.0cm,angle=270}}}
\caption{Integrated photometry for LMC Star Clusters from Persson et al.\ 
(1983).  
In both panels filled circles are confirmed old (age $>$ 10 Gyr) clusters.  
It is clear that these diagrams allow the separation of old clusters from 
intermediate-age clusters. }
\end{figure}

The field population of the LMC is known to contain RR Lyrae
stars and thus it does contain a component which is old 
(age $\geq$ 10 Gyr).  However, establishing the exact age of this old field
population is by no means simple.  Alcock et al.\ (1996, 2000), based
principally on the properties of a large sample of LMC field RR Lyraes,
have argued that the old horizontal branch in the LMC has a predominantly
red morphology.  For a mean metal abundance $<$[Fe/H]$>$ $\approx$ --1.6,
(Alcock et al.\ 1996)
and assuming that age is the second parameter governing horizontal
branch morphology, this would imply that the old metal-poor
field population is $\sim$1--2 Gyr younger than the old metal-poor LMC cluster
population (Alcock et al., 1996, 2000).  Certainly numerous field c-m diagram
studies have not revealed any significant numbers of
blue horizontal branch stars in any part of the LMC, though such stars are
found in the old globular clusters.  Direct confirmation via field c-m diagram 
studies of the assertion that the old field is younger than the old clusters 
in the LMC is a difficult process.  Even in the outskirts of the LMC, the field
population tends to be dominated by intermediate-age stars, so that separating
out and dating the old population from the range of ages and metallicities
present is far from straightforward.  However, using HST/WFPC2 based c-m 
diagrams, Walker et al.\ (1999) have found an age of $\sim$12 Gyr for the 
old field stars in an outer LMC field near the cluster NGC 2257.  They see 
little need for any additional older ($\sim$15 Gyr) component and thus their
results tend to support the ``old field is younger than the old clusters in
the LMC'' assertion.

In summary, it appears that for these (very) old ages, the integrated cluster results agree with those for the individual
clusters and for the field, though the level of precision inherent in this
statement is considerably less than for younger ages.
 
\subsubsection{2.2.2 The intermediate-age population.}

The LMC cluster system is famous for its ``age (and metallicity) gap''.
This is illustrated in Fig.\ 3 -- the only cluster known to have an age
between the age of the old globular clusters and $\sim$3 Gyr is the 
sparse cluster ESO121-SC03.  New VLT/UVES spectroscopy of
giants in this cluster has confirmed its metal abundance as [Fe/H] $\approx$
--0.9 (Hill et al.\ 2000).  
The Girardi et al.\ (1995) analysis of the $UBV$ integrated colours of the
LMC clusters not only reveals a dearth of clusters in the LMC with ages in 
excess of $\sim$3 Gyr, but also suggests that the period between $\sim$700 Myr
and $\sim$3 Gyr was an era with an enhanced cluster formation rate
(cf.\ Figs.\ 12 \& 15 of Girardi et al.\ 1995).  So once again the inferences
from the integrated cluster data are consistent with those derived from
individual cluster c-m diagram studies.

\begin{figure}[t]
\centerline{\vbox{
\psfig{figure=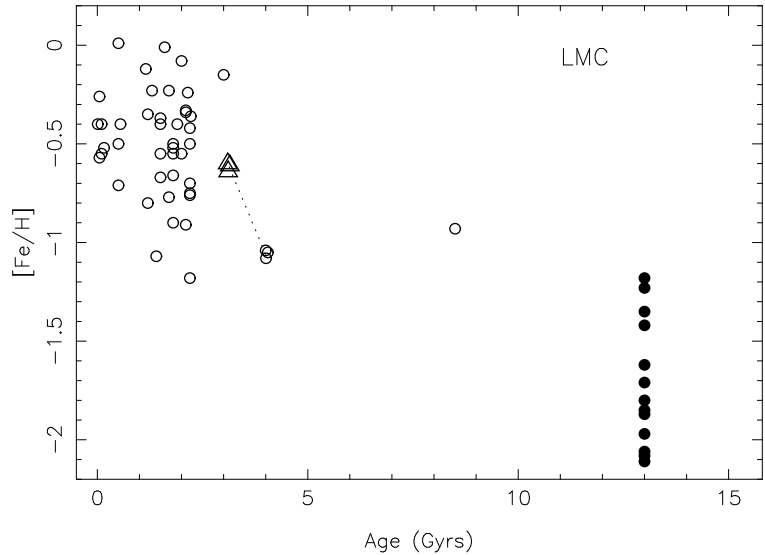,width=10.0cm,angle=0}}}
\caption{An Age-Metallicity relation for LMC star clusters.  The 13 old LMC
globular clusters have been assumed to be 13 Gyr in age.  Open symbols 
represent data from Geisler et al.\ (1997), Bica et al.\ (1998) and
Dirsch et al.\ (2000).  The triangles represent the new results of Rich et 
al.\ (2001) for the clusters NGC 2121, SL663 and NGC 2155.  The dotted line
connects these values with the older results of Sarajedini (1998) for the
same clusters.  The isolated point at age 8.5 Gyr is the cluster 
ESO121-SC03. }

\end{figure}

\subsubsection{2.2.3 The cluster ``age-gap''.}

The observed age distribution of LMC clusters with intermediate-age has 
undoubtedly been modified by the processes that lead to
cluster disruption.  However, cluster disruption does not seem to be a 
plausible complete explanation for the age-gap.  There are two 
reasons for this.  First, given that the amount of evolutionary fading 
between $\sim$3 and 12 Gyr is relatively small ($<$1 mag), it is possible
to infer from the left panel of Fig.\ 1 that at least some of the LMC
intermediate-age clusters have comparable masses to the old globular clusters.
They are therefore likely to survive for at least a few more Gyr, and hence, by
inference, if clusters older than 3 Gyr had been formed in similar numbers
they should have been discovered by now.  Certainly clusters with ages 
exceeding 3 Gyr are known in the SMC (found without any special search 
techniques), while targeted searches for age $>$ 3 Gyr clusters in the LMC 
have been notably
unsuccessful (e.g.\ Geisler et al.\ 1997).  Second, the number of 1--3 Gyr
clusters in the LMC is comparatively large (at least 40 known) and the
cutoff in numbers at $\sim$3 Gyr is relatively sharp.  It strains credulity to 
suggest the abrupt decrease in numbers at this particular age is solely a
disruption effect.
Hence, the inference from Fig.\ 3 is that the cluster formation rate
was relatively low, perhaps even zero, for ages between $\sim$3 Gyr and the
formation epoch of the LMC Galactic halo globular cluster analogues. 
The question then is, what does the field population show -- is the field
star formation rate in the $\sim$3--10 Gyr interval also relatively low (or
zero) as the cluster data appear to imply?

There have been a number of studies recently which bear on this question.
Those of particular relevance are based on HST/WFPC2 images which provide
accurate data at faint magnitudes.  Olsen (1999) has studied the star
formation history in five fields (near old globular clusters) four of which
lie in the Bar of the LMC and one in the LMC disk.  Similarly, Holtzman et al.\
(1999) have studied two outer disk and one Bar field (see also Geha et al.\
1998).  The data for the Bar fields of both Olsen (1999) and Holtzman et al.\
(1999) show a marked (factor of $\sim$3) increase in the star formation rate 
(SFR) at $\sim$5--6 Gyr compared to earlier epochs, but with no particular
evidence for any additional increase in the SFR rate at $\sim$3 Gyr.  
Nevertheless, the SFR in these fields prior
to $\sim$5--6 Gyr is evidently not zero.  The results for the disk fields
are similar -- there is no obvious increase in the SFR at $\sim$3 Gyr and,
while the SFR at epochs earlier than $\sim$5 Gyr is evidently lower than
for later epochs, it is clearly non-zero.  Consequently, in contrast to what
one might have been tempted to conclude from the existence of the cluster 
age-gap, the LMC field studies show that a significant fraction of the LMC 
field population is older than $\sim$3 Gyr.  

Clearly then we have a case where inferences from the integrated
and individual-star cluster studies give an incomplete picture of the 
star formation history of the underlying galaxy.  This outcome 
should be kept in mind when analyzing the data for more distant systems:
the properties of the cluster population may not be revealing the full 
history of the parent galaxy.  

As regards the origin of the LMC cluster age gap, an explanation may lie in 
the work of Larsen \& Richtler (2000) and Larsen (these proceedings).  
Larsen \& Richtler (2000) show that the formation of massive 
star clusters is closely linked to star formation activity, and is favoured 
at higher star formation rates.  The absolute magnitude of the brightest
(young) massive star cluster also correlates with the star formation rate
in the underlying galaxy (Larsen, these proceedings).  Hence, in the LMC,
it may be that the apparently lower overall star formation rate at ages
beyond 5 Gyr meant that not only were fewer clusters formed, but also that the
clusters which were formed were less massive and thus more readily disrupted.
Such a combination of effects could readily explain the existence of the 
cluster age-gap.
Admittedly there is a difference between the onset of increased star 
formation in the LMC Bar at $\sim$5 Gyr, and the edge of the cluster-age gap 
at $\sim$3 Gyr.  The majority of the 1--3 Gyr old clusters studied, however,
lie in the disk of the LMC rather than the Bar, and it may be that the 
epoch of increased star formation occurred later in disk.
For example, the Geha et al.\ (1998) study finds that the increase in the
star formation rate in the outer disk of the LMC occured only about 2 Gyr ago.

\section{The Small Magellanic Cloud}

The cluster system of the SMC is less numerous than that of the LMC and,
in general, has been less well studied.  Nevertheless, there are some
obvious differences between the SMC cluster system and that of the LMC\@.  In
particular, the SMC lacks any true ``old, metal-poor'' analogues to the globular
clusters of the LMC and of the Galactic halo.  The oldest SMC cluster
(NGC 121) is $\sim$2 Gyr younger than the LMC and Galactic halo 
globular clusters (e.g.\ Shara et al.\ 1998).  Further, with an abundance of 
[Fe/H] $\approx$ --1.45, it is not notably metal-poor.  Also in
contrast to the LMC, the SMC cluster system shows no indication of any 
``age gap''.  Indeed, there are numerous SMC clusters with ages between
4 and 10 Gyr, which have no counterparts in the LMC.

\begin{figure}[t]
\centerline{\vbox{
\psfig{figure=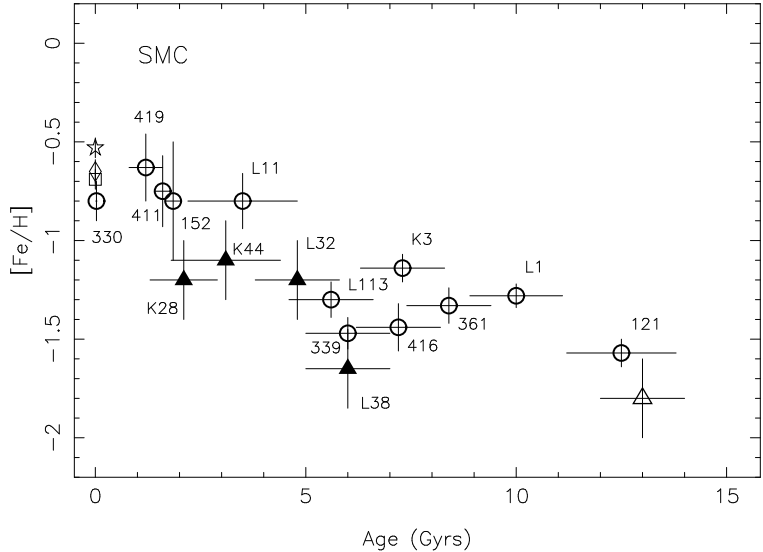,width=10.0cm,angle=0}}}
\caption{The Age-Metallicity relation for SMC star clusters.  The four 
clusters studied by Piatti et al.\ (2001; these proceedings) are plotted as
filled triangles; other clusters are plotted as open circles.  Each cluster
is identified by name.  The open triangle represents SMC field RR Lyrae stars
and the open diamond, square and star symbols are different estimates of
the present-day field star abundance (see Da~Costa \& Hatzidimitriou 1998 for
details).  Rich et al.\ (2000) have suggested on the basis of HST/WFPC2 
snapshot images
that NGC 152 and 411, and NGC 339, 416, Kron 3 and NGC 361 are in fact coeval. }

\end{figure}

Fig.\ 4 shows the age-metallicity relation for the SMC cluster system.
This figure includes new data from Piatti et al.\ (2001; these 
proceedings) while the other cluster points are drawn from Da~Costa \& 
Hatzidimitriou (1998), Mighell et al.\ (1998), de Freitas Pacheco et al.\
(1998), Rich et al.\ (2000) and earlier data.  The ``morphology'' of this
SMC cluster age-metallicity relation could be considered to have three 
components:
(i) an initial (ages $\ge$ $\sim$10 Gyr) enrichment up to [Fe/H] $\approx$ 
--1.5; (ii) a long ($\sim$3 $\leq$ age $\leq$ $\sim$10 Gyr) interval where 
cluster age and abundance are not obviously correlated; and (iii) a relatively 
abrupt increase in the cluster abundances up to approximately the present-day 
abundance of the SMC field stars.  This abundance increase seems to have 
commenced $\sim$2--3 Gyr ago.

As discussed in Da Costa (1999), this SMC cluster age-metallicity relation 
cannot be readily understood in terms of either simple chemical models 
(cf.\ Da~Costa \& Hatzidimitriou 1998), or by ``bursting models'' such as that 
of Pagel \& Tautvaisiene (1998).
The latter models rely on a very low star formation rate in the 4--12 Gyr
interval in order to reproduce the flat part of the age-metallicity relation.
However, this assumed very low star formation rate is completely 
inconsistent with the existence of many rich star clusters that have
formed during this epoch.  The most likely explanation for the flat section 
of the relation is the infall of significant amounts of relatively unenriched 
gas.  Further, Fig.\ 4 indicates that over much of the SMC's history there are 
real cluster-to-cluster abundance differences.  For example, the $\sim$0.3 dex abundance difference between Kron 3 and NGC 339 is seen in 
both spectroscopic and photometric data.  Dwarf galaxies like the SMC, 
provided the star formation rate is relatively constant and the infall of external material insignificant, are usually assumed to be chemically homogeneous, as the timescale to thoroughly mix the interstellar medium over 
dwarf-galaxy-wide scales is considerably shorter than a Hubble time (cf.\
Da~Costa \& Hatzidimitriou 1998, Da~Costa 1999).  Apparently this is not 
the case in the SMC and the explanation may again lie with infall.

As regards the integrated properties of SMC clusters, the less numerous
cluster population and the lack of a study comparable to that of Bica et al.\
(1996) mean that only general conclusions can be drawn.  Based on the data
of van den Bergh (1981), the equivalent diagrams for the SMC cluster system 
are broadly similar to those Fig.\ 1\@.  In particular, it is evident that cluster
formation has been on-going and that a similar dichotomy into ``blue'' and 
``red'' clusters at $(B-V)$ $\approx$ 0.50 exists.  The data of Persson 
et al.\ (1983) show that the younger (i.e.\ age $\sim$ 1--3 Gyr) 
intermediate-age clusters NGC 152, 411 and 419 can be readily separated from 
the older (age $\geq$ $\sim$6 Gyr) intermediate age-clusters such as NGC 339 
and Kron~3, by their red $(J-K)_{0}$ colours in the ($(J-K)_{0}, (B-V)_{0}$) 
diagram (cf.\ Fig.\ 2).  Thus it is possible to infer an extended age 
range among the SMC ``red'' clusters from their integrated properties.  
However, the ($(J-K)_{0}, (B-V)_{0}$) diagram does not separate clusters 
like NGC 339 and Kron 3 from older clusters such as NGC 121, or from 
the LMC analogues of the Galactic halo globular clusters.  The upper-AGB 
stars in the $\sim$6--10 Gyr old clusters are simply not luminous enough 
to significantly 
redden the near-IR colours relative to those for older clusters.  This result
should be kept in mind when dealing with integrated cluster colours --
it is often difficult to discern whether a particular cluster is 
$\sim$6--10 Gyr old, or $\sim$12--15 Gyr old (i.e.\ age comparable to the 
Galactic halo globular clusters) from integrated colours alone, even if
near-IR colours are also available.

As regards the SMC field population, there is very little published data
which reaches faint limiting magnitudes.  Suntzeff et al.\ (1999) report that
in a region near NGC 121, active star formation seems to have continued from 
an age approximately that of the Galactic halo globular clusters until about
6 Gyr ago, when the star formation rate decreased substantially.  
Unfortunately, given the limited data, it is not really possible to assess 
the extent to which the field star formation history and the cluster
formation history in the SMC agree, or disagree, though the clear inference 
from the discussion in Sect.\ 2.2.3 is that field star formation rate in the
SMC should exceed that in the LMC in the $\sim$5--9 Gyr interval.
Much further work remains to be done in this area.

\section{Summary}

Clearly we are still some way from a complete understanding of the 
histories of the Magellanic Clouds, our nearest galactic neighbours.  
Neverthess, in the context of this meeting, it is possible to conclude that:
first, the integrated properties of the cluster systems do give a reasonably
accurate representation of the overall characteristics in comparison with
those inferred from individual cluster studies.  However, the inferences 
become less precise
with increasing age, especially beyond a few Gyr.  Second, the LMC cluster
results suggest that one should be careful when interpreting cluster
formation histories (even after allowing for disruption processes) in terms 
of overall star formation rates for the underlying galaxy.  An absence of
clusters (not related to disruptive processes) does not necessarily mean an
absence of star formation.

\section*{Discussion}
\noindent {\it Kennicutt:\, } What are the current results on the age 
distribution in the LMC Bar, and its association (if any) with the onset
of cluster formation 3 Gyr ago?

\noindent {\it Da Costa:\, } If you combine the data for Olsen's four ``Bar'' 
fields, then the average star formation rate in the 1--3 Gyr interval is 
approximately a factor of two larger than it was over the 3--9 Gyr interval.  
It is also a factor of $\sim$4 larger than the average SFR in the 5--9 Gyr 
interval.  So the Bar is dominated by stars younger than about 5 Gyr.  The 
Holtzmann et al.\ (1999) data suggest similar results.  The increased SFR 
in the Bar seems to have commenced somewhat earlier than the increase in 
the rate of massive cluster formation, though I suspect this difference may be 
related to the fact that most of the 1--3 Gyr old clusters are in the LMC disk 
not the Bar.

\noindent {\it Frogel:\, } 1.\  Is there a difference in [Fe/H] distribution
of LMC and Milky Way RR Lyrae that could account for the inferred age
difference?  2.\ If you look at specific cluster frequency per number of
field stars, is the ``age-gap'' still really a gap?

\noindent {\it Da Costa:\, } 1.\ Alcock et al.\ (1996) give an [Fe/H]
distribution based on spectra of only 15 LMC field RR Lyraes.  This 
distribution, which peaks at [Fe/H] $\approx$ --1.6 with a prominent tail
to lower abundances and an abrupt cutoff at [Fe/H] $\approx$ --1.0, is
broadly similar to the [Fe/H] distribution for Galactic halo field RR Lyrae.
But a larger sample of LMC field RR Lyrae abundances is needed to give
statistical weight to this similarity (or any difference).  
2.\ If the 
specific frequency of clusters is constant, then since the overall (Bar and
disk) star formation rate in the 1--3 Gyr interval is $\sim$3 times that
in the 3--9 Gyr interval, we would expect roughly equal numbers of 1--3 Gyr 
and 3--9 Gyr old clusters if we ignore fading and disruption.  Fading 
($\leq$1 mag from 3 to 9 Gyr) will reduce the relative number of older
clusters in any magnitude limited (i.e.\ observed) sample and undoubtedly
a number of (potential) older clusters, particulary those of lower mass, will
have been disrupted.  Yet we know of at least 40 clusters whose ages,
determined from main sequence turnoff photometry, lie in the 1--3 Gyr range  
while, despite significant searches (e.g.\ Geisler et al.\ 1997), 
there is only {\it one} LMC cluster known whose age exceeds 3 Gyr and which 
is younger than the LMC Galactic globular cluster analogues (ESO 121-SC03).  
It seems to me highly unlikely that this disparity results solely from a 
combination of fading and disruption.  The ``age-gap'' is a real phenomenon.

\noindent {\it Lamers:\, } Is it possible that the presence of an age gap
for the LMC clusters, and the lack of an age gap for the field stars, is due
to the fact that most clusters with age $>$ 3 Gyrs have dispersed?  In my
(poster) paper I show that the disruption time depends on the initial cluster
mass as $t_{dis}$ $\sim$ M$^{+0.6}$, so the lower mass clusters may have
dispersed and only the most massive ones survived.

\noindent {\it Da Costa:\, } Disruption of LMC clusters must play some role 
in generating the ``age-gap''.  However, I do
not believe that disruption processes can be entirely responsible for the 
relatively abrupt cutoff in the cluster age distribution at $\sim$3 Gyr.  
There have been a number of extensive searches for LMC clusters with ages in 
excess of $\sim$3 Gyr and all have been singularly unsuccessful.  Further, a 
number of current 1--3 Gyr old LMC clusters have luminosities
that indicate initial masses exceeding 10$^{4}$ M$_{\sun}$ and thus, using
the results from your poster, disruption times in excess of 5 Gyr.  It is then
difficult to see, if LMC clusters had formed at a uniform rate, why the age
gap would exist. 

\end{document}